# Magnetic-field-induced nontrivial electronic state in the Kondo-lattice semimetal CeSb

Y. Fang[1,*], F. Tang[1], Y. R. Ruan[2], J. M. Zhang[2,*], H. Zhang[3], H. Gu[1], W. Y. Zhao[4], Z. D. Han[1], W. Tian[5], B. Qian[1], X. F. Jiang[1], X. M. Zhang[6], and X. Ke[3,*]

[1]Jiangsu Laboratory of Advanced Functional Materials, Department of Physics, Changshu Institute of Technology, Changshu 215500, China

[2]Fujian Provincial Key Laboratory of Quantum Manipulation and New Energy Materials, College of Physics and Energy, Fujian Normal University, Fuzhou 350117, China.

[3]Department of Physics and Astronomy, Michigan State University, East Lansing, Michigan 48824, USA

[4]ISEM, Innovation Campus, University of Wollongong, Wollongong, NSW 2500, Australia

[5] Neutron Scattering Division, Oak Ridge National Laboratory, Oak Ridge, Tennessee 37831, USA

[6]School of Materials Science and Engineering, Hebei University of Technology, Tianjin 300130, China

[*]Corresponding address: fangyong@cslg.edu.cn (Y. Fang), jmzhang@fjnu.edu.cn (J. M. Zhang), and kexiangl@msu.edu (X. Ke)

**Abstract**

Synergic effect of electronic correlation and spin-orbit coupling is an emerging topic in topological materials. Central to this rapidly developing area are the prototypes of strongly correlated heavy-fermion systems. Recently, some Ce-based compounds are proposed to host intriguing topological nature, among which the electronic properties of CeSb are still under debate. In this paper, we report a comprehensive study combining magnetic and electronic transport measurements, and electronic band structure calculations of this compound to identify its topological nature. Quantum oscillations are clearly observed in both magnetization and magnetoresistance at high fields, from which one pocket with a nontrivial Berry phase is recognized. Angular-dependent magnetoresistance shows that this pocket is elongated in nature and corresponds to the electron pocket as observed in LaBi. Nontrivial electronic structure of CeSb is further confirmed by first-principle calculations, which arises from spin splitting in the fully polarized ferromagnetic state. These features indicate that magnetic-field can induce nontrivial topological electronic states in this prototypical Kondo semimetal.

**Introduction**

Correlated electronic materials, in which the competing interactions are nearly degenerate in energy scale, attract considerable attention due to the emergence of exotic physical phenomena [1-3]. An excellent example is the lanthanides (Ce, Sm, Yb, etc)-based intermetallic compounds, where the hybridization between localized 4$f$-electrons and conduction electrons produces a broad range of electronic states [4-6]. Of particular interest are Kondo insulating and Kondo semimetallic states. Kondo insulators, such as CeRu$_4$Sn$_6$ and Ce$_3$Bi$_4$Pt$_3$, behave as correlated metals at high temperature due to the weak many-body interactions between local spins of lanthanide and conduction electrons, but they show small (~ 10 meV) electronic band gaps featured by rapid increment in resistivity at low temperature [7, 8]. In contrast, Kondo semimetals, such as CeNiSn, and CeNi$_{2-\delta}$As$_2$ [9, 10], exhibit metallic behaviors throughout the whole temperature range.

Interestingly, in the Ce-based systems, particularly those materials with low carrier density, the combination of strong spin-orbit coupling, opposite parities between 4$f$ and 5$d$ bands, and hybridization-induced gaps, may give rise to nontrivial topological phases with strong electron correlation [11-13]. For instance, topologically protected hourglass-type band structures with Möbius surface states in CeRhSb have been predicted by density functional theory calculations and verified by photoemission spectroscopy [14-15]. And CeRu$_4$Sn$_6$ is predicted to be a Weyl semimetal with broken inversion symmetry at the ground state [16]. Nevertheless, experimental investigation of topological electronic properties of correlated *f*-electron systems is still rare.

CeSb, a binary Kondo system, is proposed to be another member of promising topological semimetal candidate [17]. It crystallizes in a simple rock-salt structure with the space group symmetry $Fm\bar{3}m$ at room temperature and exhibits unusual magnetism with complicated phase diagram [18]. Previous neutron scattering studies found a series of commensurate phases that are composed of various stacking sequences of ferromagnetic layers intermixed with paramagnetic ones along the c-axis in this compound [19]. Similar to other well-known topological semimetals, extremely large magnetoresistance, which generally arises from the electron-hole (*e-h*) compensation [20-22], is observed in CeSb [17]. However, its topological nature and the stability of surface states remain controversial. Early angle-resolved photoemission spectroscopy (ARPES) experiments revealed two anisotropic and temperature-independent Dirac-cone-like bands in CeSb, suggesting that it is a nontrivial topological semimetal [24]. And Guo *et al* reported that in the nonmagnetic state there are two band inversions with very small splitting below the Fermi level, of which the calculated $Z_2$ classification is (1;000), implying the nontrivial band structure of this compound as well [17]. In contrast, soft-x-ray ARPES measurements on CeSb suggested that the unusual Dirac-cone-like energy dispersions were not of topological origin but could be derived from significant $k_z$ broadening instead [25, 26]. Intriguingly, its sister compound CeBi is found to be a topological semimetal, and it was suggested that CeSb is on the border between topological trivial and topological nontrivial semimetals [25].

This brings about an interesting question: could the electronic band structure of CeSb and its topological nature be altered by magnetic field? Recently, it has been reported that a half Heusler compound GdPtBi exhibits linear electronic band crossing and Weyl points around the Fermi level in the presence of magnetic field, which is further evidenced by negative longitudinal magnetoresistance (chiral anomaly) [27]. Similarly, CeSb also shows pronounced angle-sensitive negative magnetoresistance, which is ascribed to charge pumping from one Weyl point to the other [17]. This is supported by the electronic structure calculations, where band inversions and non-zero integer Chern number are obtained in the fully polarized ferromagnetic state [17]. Nevertheless, the ARPES measurements and band structure calculations reveal that the band inversions are localized at more than 300 meV below the Fermi level [17, 24-26], which should have weak effect on transport. Thus, whether the negative magnetoresistance in CeSb originates from chiral anomalies or not is still an open question, and the nontrivial electronic features of CeSb in the fully polarized ferromagnetic state demands further investigations. In general, although high-resolution ARPES method is the best choice to clarify the band structure and determine the topological nature of materials, this technique does not work in the presence of magnetic field. Alternatively, the electronic state hosting topologically nontrivial or trivial features can also be indirectly probed by the Berry phase, which can be directly obtained from the quantum oscillation analysis [28, 29].

To gain more insight into the physical properties, particularly the topological nature of electronic states of CeSb, we have grown high-quality single crystals and performed neutron diffraction, specific heat, magnetization and electronic transport measurements. Clear quantum oscillation emerges both in the magnetization and magnetoresistance at high fields, from which the nontrivial Berry phase is extracted. This suggests that CeSb is indeed a topological semimetal candidate in its fully polarized ferromagnetic state. The nontrivial nature of this compound is further corroborated by the electronic band structure calculations, where two valence bands and two conduction bands cross between Γ and X point in the Brillouin zone, forming three inverted bands with small gaps and a band crossing point.

**Experimental details**

Single crystals of CeSb were grown by molten salt method employing tin as flux [17]. Single crystal neutron diffraction experiments were carried out using the HB-1A triple-axis spectrometer at High Flux Isotope Reactor (HFIR) in Oak Ridge National Laboratory. The energy of incident neutrons of HB-1A was fixed as $E_i = 14.6$ meV. The samples were oriented in (*H* 0 *L*) and (*H H L*) scattering planes, where *H*, *K* and *L* are in reciprocal lattice units (*r. l. u.*) $2\pi/a$, $2\pi/b$ and $2\pi/c$ with $a=b=c=6.416$ Å at room temperature. Samples were mounted in an aluminum can and cooled using a closed-cycle Helium refrigerator down to 5 K. The magnetization measurements were performed on a Quantum-Design Physical Property Measurement System (PPMS) equipped with a vibrating sample magnetometer (VSM) option. The resistivity was determined by the conventional four-probe method using the Quantum-Design PPMS cryostat equipped with rotator option. Before measurement, a platelet of sample with face perpendicular to the [001] direction was polished from a faceted crystal. Good quality electrical contacts were made with gold wires (25 μm) glued on the crystal surface using silver paste (Dupont 4929N).

**Theoretical details**

The electronic band structure of CeSb in the fully polarized ferromagnetic state were examined by the density functional theory (DFT) calculations, as implemented in Vienna *ab initio* simulation package (VASP) [30, 31]. The stacking fault energy calculations were carried out using the projected augmented wave (PAW) functions and generalized gradient approximation (GGA) with electron exchange-correlation described by Perdew-Burke-Ernzerhof (PBE) parameterization [32, 33]. A 14×14×14 Monkhorst-Pack *k*-point mesh was employed with a cutoff energy of 450 eV. The GGA+U was adopted in the Dudarev's approach for the electronic structure calculation [34], where the on-site Hubbard *U* for the correlated Ce-4*f* orbitals are $U_f=$ 6.7eV and $J_f=0.66$ eV [17]. Here, we adopted the Ce atoms with periodic boundary conditions arranged in ferromagnetic ordering along the [001] direction. The magnetic moment of Ce was calculated to be about 1 $u_B$. During the calculations, spin-orbit coupling effect was taken into account.

**Results and discussion**

Figure 1(a) shows the specific heat versus temperature for CeSb, in which seven anomalies are clearly observed between 2 and 20 K, indicating that there are successive phase transitions upon cooling the compound from high temperature. Figure 1(b) shows the complementary magnetic susceptibility measured at a field of 0.05 T along the [001] direction. Seven anomalies are also found in the magnetic susceptibility $\chi$ and its derivative $\partial(\chi \cdot T)/\partial T$ at essentially the same temperatures as observed in specific heat, suggesting that these anomalies are associated with magnetic phase transitions. Previous neutron diffraction studies on CeSb have reported different numbers of phases, such as 3, 5 or 6 as mentioned in Ref. 35 and 36, which were ascribed to the difference in crystal quality including purity and crystallinity. Our single crystal neutron diffraction measurements, as presented by the (2 0 $\delta$)-$T$ contour map in Fig. 1(c), have shown both similarity to and difference from the earlier studies [35, 36]. In agreement with previous reports, the propagation wave vector of CeSb is found to be (0 0 $\delta$). $\delta$ can be expressed using a general formula as $\delta=n/(2n-1)$ with $n$ being an integer, exhibiting strong temperature dependence, as highlighted by the circular dots in Fig. 1(c). By comparing Fig. 1(c) with the specific heat and magnetic susceptibility data, we can see that neutron diffraction measurements have clearly revealed the first 6 magnetic phases except the one at $T_7$. $\delta$ = 2/3, 8/13, 4/7, 5/9, and 6/11 for the first 5 phases at higher temperatures, consistent with those reported in Ref. 35 and 36. The corresponding $n$ values are 2, 7, 4, 5, 6, respectively. For these phases, the associated magnetic structures were denoted as AFP$_n$ [18], where there exists a nonmagnetic layer every (2$n$-1) ferromagnetic layers along the c-axis. At $T=T_6$=9.3 K, we find that CeSb undergoes a transition to a phase with a propagation wavevector of (0 0 0.5) which corresponds to an AFF phase with up-up-down-down type spin structure along the c-axis and without any nonmagnetic layers [35]. At $T=T_7$, no change of the propagation wavevector is observed. This, combined with the sharp increase in magnetic susceptibility at $T_7$, suggests the likelihood of having a canted AFF phase at lower temperatures. This is also consistent with the very small anomaly observed in specific heat data at $T_7$. These features are in contrast with an earlier literature which reported a change of the wavevector to (0 0 0.5) at $T_7$ but not at $T_6$ [36]. Figure 1(d) shows rocking curve scans of (2 0 2+δ) and (0 0 2+δ) magnetic Bragg reflections with δ=0.5 measured at $T$=5.6 K and δ=4/7 measured at $T$=15.3 K. The larger intensity observed for the (2 0 2+δ) Bragg peak compared to the background-like feature for the (0 0 2+δ) indicates that spins are aligned along the c-axis at both low and high temperatures.

Figure 2(a) presents the isothermal magnetization measured with fields applied along the [001] direction from 2 K to 28 K with a temperature interval of 1 K, where several field-induced transitions can be clearly observed. At 2 K, it undergoes approximately five magnetic phases evolving from four successive antiferromagnetic states eventually to a ferromagnetic state with two abrupt jumps of magnetization at the critical fields $H_{C1}$~1.9 T and $H_{C2}$~3.7 T. Between these two magnetic fields, a magnetization plateau $M/M_s \approx 0.33$, which is quite close to the value of 1/3, is observed. The step-like magnetization could be ascribed to flips of spins for the one-dimensional Ising chain nature of CeSb [37]. The magnetization saturates and reaches 2.2 $\mu_B$/Ce above 3.7 T, which corresponds to the magnetic moment in the ordered state with full moment of $J$= 5/2 for Kramers ion $Ce^{3+}$ [38]. After the magnetic moments are fully polarized, clear de Haas-van Alphen (dHvA) oscillations emerge. For $T \leqslant 8$ K, the shape of magnetization curves basically remains unchanged with only the plateau regime slightly growing. At 9 K<$T$<16.8 K, there appear new field-induced magnetic phases and the saturation field gradually enhances. For instance, at 11 K there are two magnetization plateaus corresponding to 1/3 and 2/3 of the saturated magnetic moment respectively. Upon further increasing the temperature, multiphases constituted by at least 5 to 7 magnetic states are found between 11 K and 16.8 K.

To better visualize the dHvA oscillations, Fig. 2(b) shows an expanded view of the high-field region of magnetization measured with field applied along the [001] direction. The quantum oscillations, which become more obvious after subtracting the ferromagnetic background, decrease gradually upon warming the sample. Figure 2(c) presents the plots of oscillatory component $\Delta M$ versus $(\mu_0 H)^{-1}$, where one can clearly see periodic oscillations of $\Delta M$. To understand the nature of these dHvA oscillations [39, 40], fast Fourier transformation (FFT) analyses of the oscillatory magnetization is carried out in the field range between 5.5 and 9 T. As shown in Fig. 2(d), the field dependence of FFT amplitude exhibits a single frequency with $F \sim 198$ T. Using the Onsager relation $F=(\Phi_0/2\pi^2)S_F$ [40], we can extract the extremal Fermi surface cross-sectional area normal to magnetic field to be $S_F \sim 1.8 \times 10^{-2}$ Å$^{-2}$, which corresponds $\sim 1.8\%$ of the total area in first Brillouin zone. Here, $\Phi_0$ (=2.17×10$^{-15}$ $T \cdot m^{-2}$) is the magnetic flux quantum [40]. Assuming a circular cross section, a small Fermi wave vector of $\sim 7.5 \times 10^{-2}$ Å$^{-1}$ is estimated, indicating that the Fermi surface of CeSb is not point-like, which is in agreement with the energy band structure shown later on. Figure 2(e) presents the temperature dependence of FFT amplitudes. Fitting the curve using Lifshitz-Kosevich (LK) formula with a Berry phase being taken into account for a three-dimensional Dirac system [39], the effective mass $m^*$ to be of 0.23 $m_e$ can be extracted.

It is known that in a Dirac system the linear band dispersion gives rise to an extra π Berry phase [39, 40]. To determine this feature, in Fig. 2(f) we plot the Landau level fan diagram with the inverse magnetic field $1/(\mu_0 H)$ being the *x*-axis and the integer Landau level index N as the y-axis [39, 40]. Note that the valleys and peaks of oscillations are assigned with the Landau level index of *N*-3/4 and *N*-1/4, respectively, and the *N* values associated with peaks and valleys of oscillations form a straight line as a function of $1/(\mu_0 H)$, which can be fitted using the formula $N=F/(\mu_0 H)+\phi$ [41]. The inset of Fig. 2(f) shows the typical oscillation pattern at 2 K, in which the Landau index number *N* can be labelled. For quantum oscillation with single frequency, the phase shift should take the value of ±1/8 for three-dimensional Dirac fermions [42]. As shown, the obtained intercept (~ -0.11) is close to the theoretical range for three-dimensional Dirac fermions but far from that expected for a parabolic dispersion (intercept ~0.5), which clearly demonstrates the presence of three-dimensional Dirac fermions in the ferromagnetic state of CeSb [42]. Besides, for magnetic field applied along the *a* axis, an extrapolated value ~0.09 (Fig. S1 in Supplementary Materials [43]) is extracted from the Landau level fan diagram, reinforcing the evidence that ferromagnetic CeSb is topological in the electronic band structure.

Figure 3(a) shows the magnetoresistance as a function of magnetic field below 10 K. The inset of Fig. 3(a) illustrates the experimental configuration. One can see extremely large positive magnetoresistance at low temperatures, which does not saturate up to 9 T, consistent with the previous reports [17, 18]. In addition, kinks around 3.8 T are observed, corresponding to the ferromagnetic phase transition seen in Fig. 2(a). Interestingly, weak but discernable Shubnikov-de Haas (SdH) quantum oscillations are also observed above 3.8 T. The oscillatory component of magnetoresistance at various temperatures is obtained by subtracting the smooth magnetoresistance background and plotted as a function of the inverse magnetic field $1/(\mu_0 H)$ in Fig. 3(b). The FFT spectra of SdH oscillations, as shown in the inset of Fig. 3(c), give only a principal frequency at $F\sim 190$ T, which is close to the dHvA oscillation frequency shown in Fig. 2(d). From the temperature dependence of oscillation amplitudes plotted in Fig. 3(c), the effective mass, 0.25 $m_e$, is obtained using the LK theory [44], which again agrees with the dHvA measurements presented in Fig. 2(e). Since the transverse resistivity $\rho_{yx}$ is far less than the longitudinal resistivity $\rho_{xx}$ in semimetals, the peaks and valley of SdH oscillations can be defined as integer indices and half integer indices respectively to determine the phase shift $\phi$ [45]. As shown in Fig. 3(d), the linear fitting of $1/(\mu_0 H)$ dependence of $N$ using $N=F/(\mu_0 H)+\phi$ gives an intercept value of -0.12, which agrees well with the value determined from dHvA measurements shown in Fig. 2(d). Furthermore, we plot the Landau level fan diagram (Fig. S2 in the Supplementary Materials [43]) based on SdH oscillation [Fig. 4(b)] with the magnetic field applied along different angles relative to the *c* axis in *ac* plane, from which it can be found that the intercepts are angular-dependent but keep their values between -1/8 and +1/8, indicating the ferromagnetic CeSb is topologically nontrivial.

To understand the Fermi surface geometry, it is necessary to measure the angular-dependent SdH. Inset of Fig. 4(a) shows the schematic diagram for magnetoresistance measurement in which the magnetic field changes from the in-plane direction to the out-of-plane direction upon rotating the sample. In Fig. 4(a), one can see that the magnetoresistance drops rapidly when the field is titled away from the *c*-axis to the *ab* plane. Such large differences between the out-of-plane and in-plane magnetoresistance have been widely observed in $WTe_2$, LaSb, $NbSb_2$, *etc* [46-48], which indicates the normal component of magnetic field governs the magnetoresistance. Furthermore, there exist SdH oscillations in the magnetoresistance measured at high magnetic fields. To investigate these features in detail, Figure 4(b) shows the extracted magnetoresistance oscillatory component after subtracting the polynomial background from the magnetoresistance signal measured at various angles. And the corresponding frequency spectra obtained via FFT analysis is presented in Fig. 4(c). One can see that the oscillation feature/frequency significantly changes, evidencing the anisotropic three-dimensional Fermi-surface pockets [21]. Specifically, the frequency *F* is 190 T at $\theta=0°$ ($\mu_0 H // c$), and it increases when the field gradually rotates towards 90° ($\mu_0 H // a$). With $\theta$ reaching 30°, an additional frequency $F'$ emerges just above *F*, and decreases upon increasing the angle further. Note that $F'$ doesn't correspond to the fundamental frequency of (001) plane but arises from the neighboring (100) plane due to the crystal symmetry. Once the field is applied along the *a*-axis ($\theta=90°$), $F'$ acts as the fundamental frequency corresponding to the (100) plane, while F of the (001) plane disappears completely. Similar features have been found in the sister compound LaBi and cubic topological insulator $SmB_6$ [21, 49]. These features suggest that the carrier pocket of CeSb hosts elongated Fermi surface along the c direction. In general, for the elongated ellipsoidal Fermi surface pocket with anisotropic three-dimensional feature, the angular dependence of cross-section area A of the pocket as well as its corresponding quantum oscillation frequency F can be described as A~F~$\frac{\pi ab}{\sqrt{sin^2\theta+\left(\frac{a^2}{b^2}\right)cos^2\theta}}$ (a and b denote the semimajor and semiminor axes of the ellipsoid, respectively) [50]. With a>b, A~F can be approximated as $\frac{\pi b^2}{cos\theta}$, which has been observed in LaSb and LaBi systems [47, 50]. Figure 4(d) plots the frequency F as a function of θ, which can be nicely fitted using F/cos(θ-nπ/2) (n=0, red solid line), showing a similar case with those in LaSb and LaBi [47, 50].

To better understand the quantum oscillation behavior of CeSb in the fully polarized ferromagnetic states and its topological nature, we have performed first-principle electronic band structure calculations taking into account the spin-orbit coupling. Figure 5 shows the band structure along high-symmetry directions of the Brillouin zone (BZ) and the corresponding Fermi surfaces. Here, only the bands around Fermi level are concerned. It is worth noting that no band crossing the Fermi level along other high-symmetry directions in the BZ can be found, except those presented in Fig. 5(a). In general, nonmagnetic RSb compounds exhibit doubly degenerate electronic bands due to the space and time inversion symmetries [51, 52]. Recently, Kuroda *et al* showed that CeSb in the paramagnetic state possesses two doubly-degenerate bands and one doubly-degenerate band crossing the Fermi level around the Γ point and the X point, respectively [25]. It was suggested that CeSb is topological trivial due to the absence of band inversion, even though it shows Dirac-cone-like energy dispersion with a small gap of ~0.1 eV [25]. Interestingly, as shown in Fig. 5(a), the two-fold degeneracy of the bands is lifted in the ferromagnetic phase due to the time-reversal symmetry breaking. Furthermore, band inversions of the Ce-5$d$ and Sb-5$p$ orbitals, as highlighted in the dashed box of Fig. 5(a), occurs along the Γ-X direction at 0.25~0.4 eV below the Fermi level, which agrees well with the prediction of Jang *et al* [4]. In addition, the $Z_2$ topological index is evaluated according to the parity criteria proposed by Fu and Kane [53], which is calculated based on the parity of all occupied states at the eight time-reversal invariant momentum points. Here, the parity of a band can be determined by a symmetry analysis of the orbitals. In the ferromagnetic state, the valence and conduction band of CeSb near X point are derived from Sb-$p$ and Ce-$d$ orbitals respectively. At the X point, the parity of Sb-$p$ band is odd, while the parity of Ce-$d$ band is even. As a result, nonzero topological invariant $Z_2$ can be obtained due to the inversion of Sb-$p$ and Ce-$d$ bands at the X point, which indicates that CeSb is a nontrivial semimetal in its ferromagnetic state.

Furthermore, the Fermi surface [Fig. 5(b)] obtained for CeSb in the ferromagnetic state is remarkably different from those of the nonmagnetic $R$Sb ($R$=rare earth) [26, 54], but similar to that of HoSb which was identified as a topologically nontrivial semimetal [55]. The electronic structure calculation reveals that CeSb has two electron-pockets and four hole-pockets in the fully polarized ferromagnetic state. The hole-pockets (β, β′, γ, γ′) are centered at the Γ points, while two orthogonally arranged ellipsoidal triple electron-pockets (α, α′) elongate along the Γ-X path. By roughly evaluating the volumes of those pockets ($V_\alpha \approx 62.74\ nm^{-3}$, $V_{\alpha'} \approx 44.24\ nm^{-3}$, $V_\beta \approx 45.79\ nm^{-3}$, $V_{\beta'} \approx 28.31\ nm^{-3}$, $V_\gamma \approx 15.42\ nm^{-3}$, $V_{\gamma'} \approx 12.07\ nm^{-3}$), we find that the corresponding carrier density ratio is $n_e/n_h \sim V_e/V_h = 1.05$, indicating that the compensated nature of CeSb still remains in the ferromagnetic state. Additionally, CeSb exhibits highly anisotropic topology of the Fermi surface, except for those two hole pockets (γ and γ′), as shown in Fig. 5(c)-(e) for the Fermi surface projection in the (001), (110) and (111) planes. Notably, the two ellipsoidal electron pockets have an aspect ratio 2.7 and 2.8 respectively, indicating that the corresponding bands are elongated in nature. By calculating the cross-sectional area of each pocket in (001) plane, we find that the obtained value for electronic pocket α is $S_F=1.73\times10^{-2}$ Å$^{-2}$ ($F$=190 T), which is smaller than that obtained from dHvA oscillation ($F$=198 T), but basically consistent with that revealed by SdH oscillation. Slight difference between the calculated and experimental frequencies can be easily found elsewhere [52]. This result suggests that the calculated Fermi surface indeed captures the true electronic state of CeSb in the ferromagnetically polarized state. It is worth noting the fact that only one oscillation frequency is observed in both dHvA and SdH data due to the weak nature of those oscillations for high frequency component in the measured temperature and field range.

**Conclusion**

    Neutron diffraction and specific heat measurements clarify that CeSb

experiences several magnetic transitions with the propagation wavevector at $T_7$ unchanging and at the ground this compound possibly shows canted AFF spin structure. With the magnetic field increasing, several field-induced phase transitions generate. In the ferromagnetic state, quantum oscillations in magnetization and magnetoresistance are clearly observed, from which the nontrivial electronic properties of CeSb is revealed. The experimental observation is corroborated by the electronic structure calculations which reveal band inversion of Ce-5$d$ and Sb-5$p$ orbitals at 0.25~0.4 eV below the Fermi level in the polarized state.

**Acknowledgments**

This work is supported by the National Natural Science Foundation of China (Grant No. 11604027, 11874113 and U1832147), Key University Science Research Project of Jiangsu Province (19KJA530003), Natural Science Foundations of Fujian Province of China (Grant No. 2017J06001), Open Fund of Fujian Provincial Key Laboratory of Quantum Manipulation and New Energy Materials (Grant No. QMNEM1903). Work at Michigan State University was supported by the National Science Foundation under Award No. DMR-1608752 and the start-up funds from Michigan State University. A portion of this research used resources at the High Flux Isotope Reactor, a DOE Office of Science User Facility operated by the Oak Ridge National Laboratory.

**Figures:**

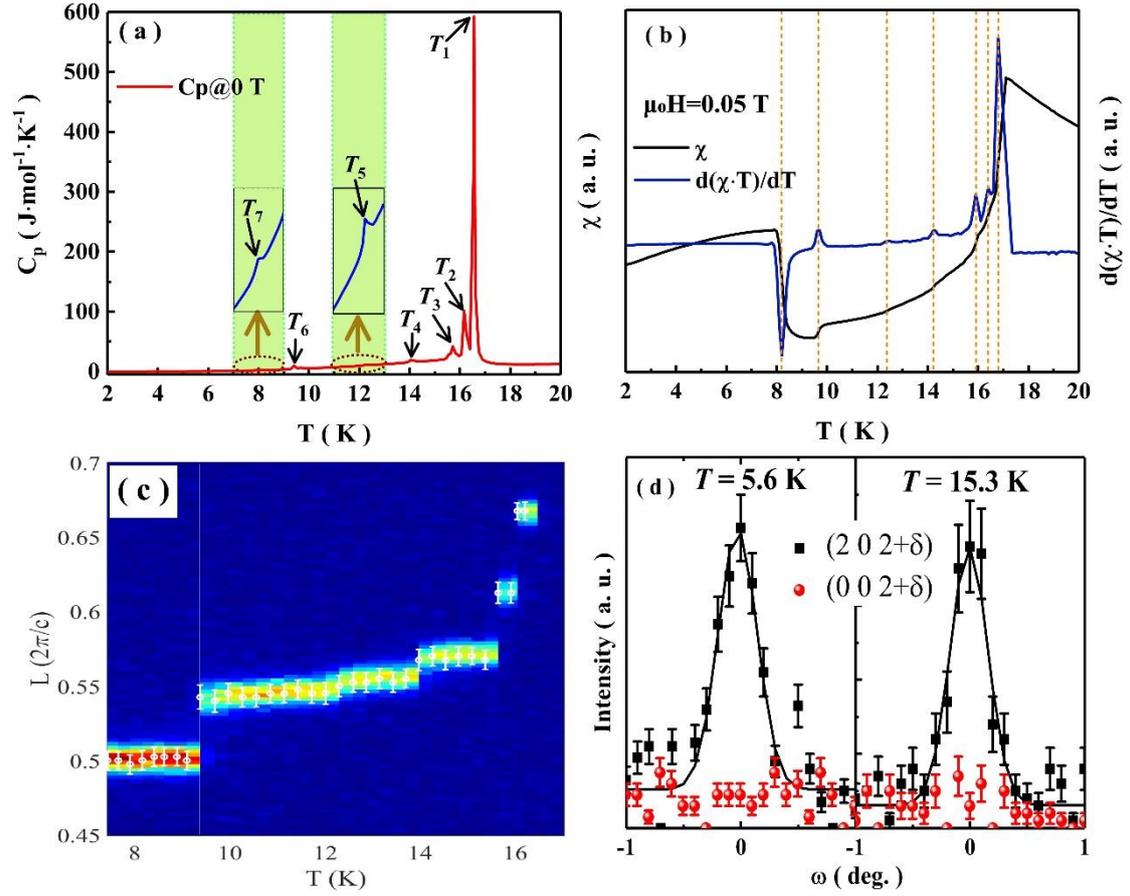

FIG. 1 (a) Specific heat as a function of temperature under zero magnetic field. (b) Temperature-dependent magnetic susceptibility and its derivative $\partial(\chi \cdot T)/\partial T$ versus temperature measured at 0.05 T. (c) Neutron contour map of (2 0 L) vs temperature. Circular dots denote L values measured at each temperature. (d) Rocking curve scans of (2 0 2+δ) and (0 0 2+δ) with δ=0.5 measured at $T$=5.6 K and δ=4/7 measured at $T$=15.3 K.

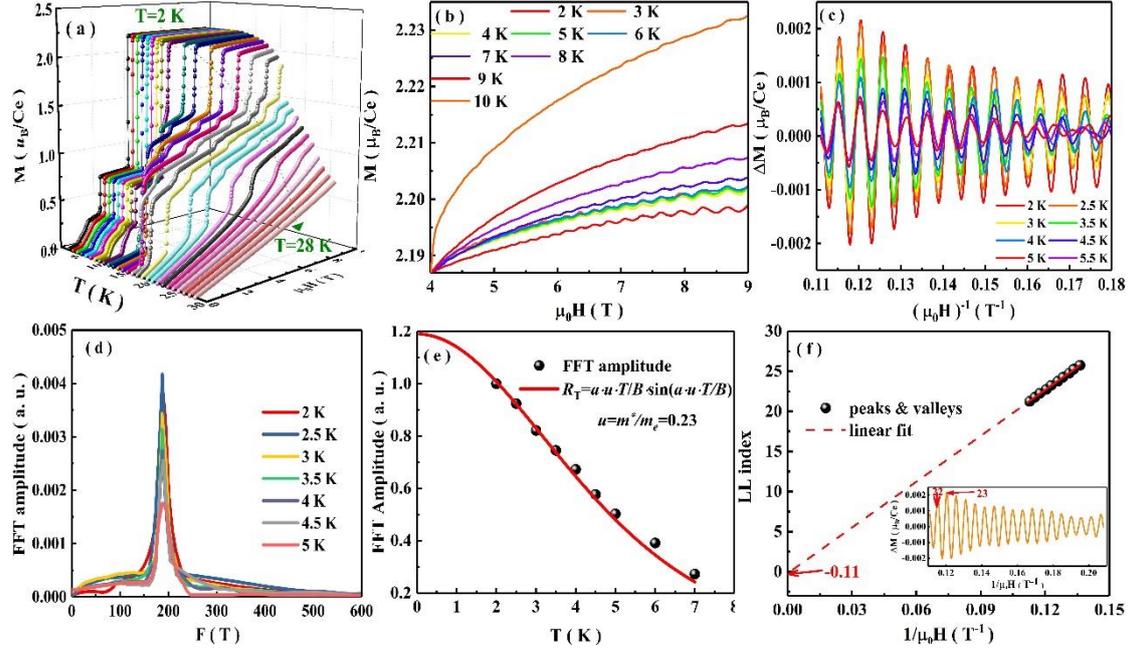

FIG. 2 (a) Magnetization versus field measured at various temperatures between 2 K and 28 K with field applied along the c-axis. (b) High-field magnetization for CeSb at different temperatures. (c) Extracted oscillatory components of magnetization at various temperatures. (d) FFT spectra of the oscillatory components shown in (c). (e) Temperature dependence of the extracted FFT amplitude. (f) Landau-level index plots *N* versus 1/($\mu_0$H).

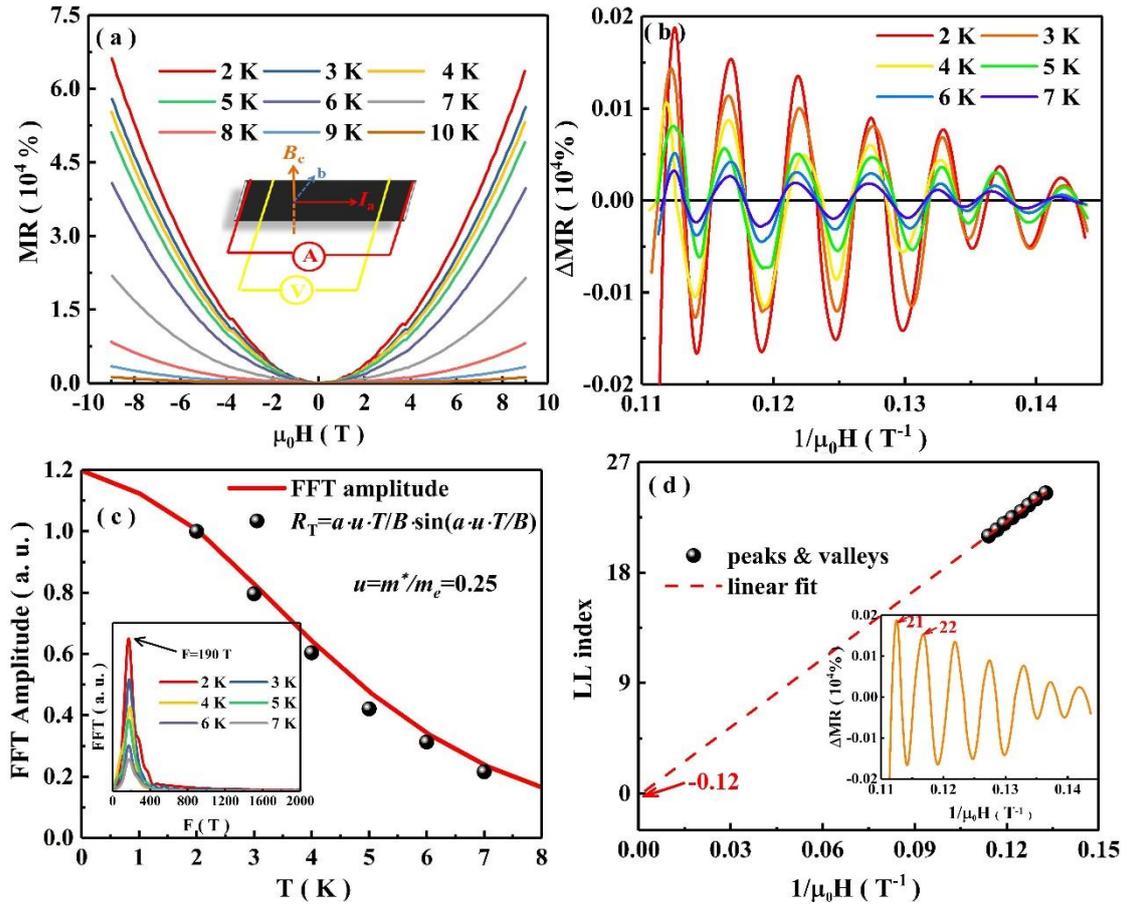

FIG. 3 (a) Magnetoresistance measured at various temperatures between 2-10 K. (b) SdH oscillatory component of resistance at different temperatures. (c) Temperature dependence of out-of-plane ($\mu_0H$ //c) FFT amplitude; the solid lines represent the LK fit for effective mass. The inset shows the FFT spectra of the oscillatory components. A frequency around 190 T is clearly observed. (d) Landau index *N* plotted against $1/(\mu_0H)$. The inset shows a typical oscillation pattern at 2 K, in which the Landau index number *N* can be labelled.

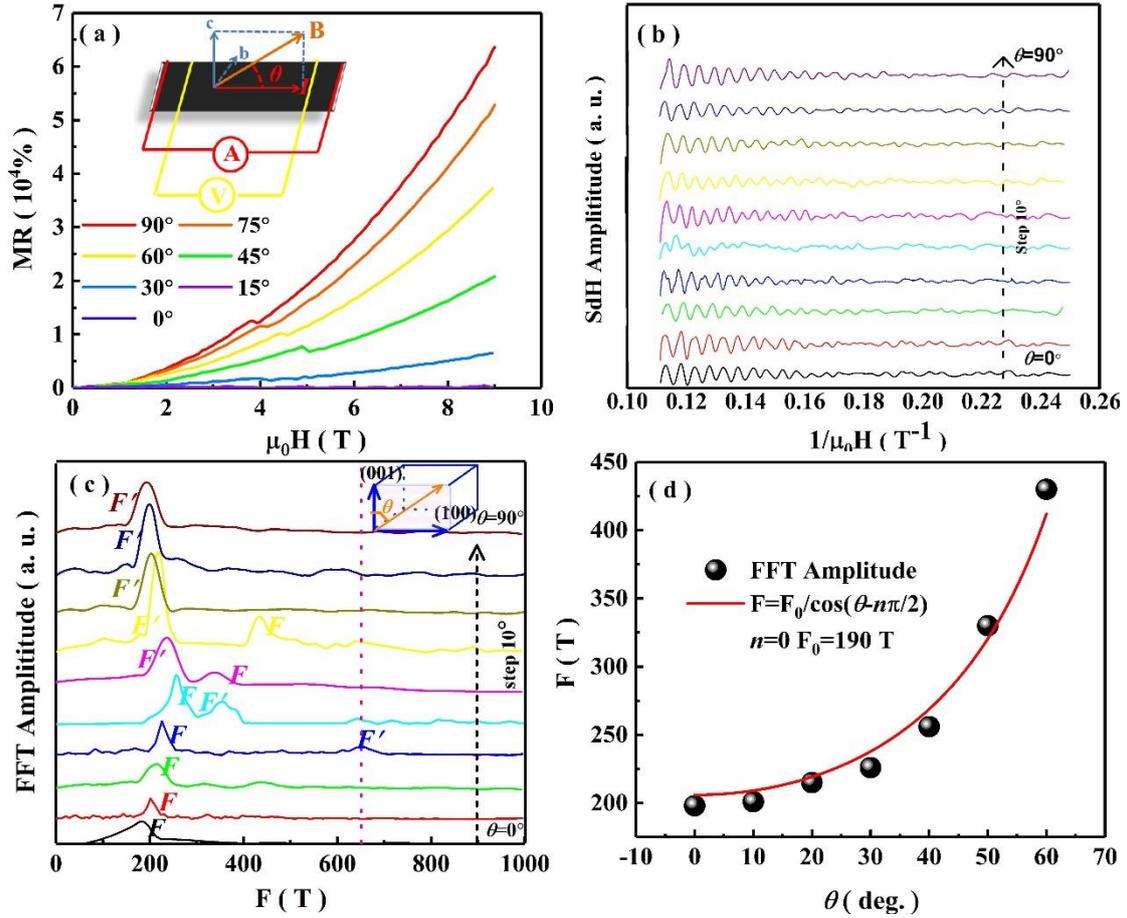

FIG. 4 (a) Magnetic field dependence of magnetoresistance at 2 K measured at various angles. Inset shows the sketch of the measurement configurations, where the magnetic field rotates in *ac* plane. (b) The SdH oscillation amplitude versus $1/(\mu_0 H)$ after subtracting the magnetoresistance background at different tilting angles. (c) FFT spectra of the SdH quantum oscillations with field rotated in *ac* plane. Inset in shows the definition of field orientation angle $\theta$. (d) Angular dependence of the SdH oscillation frequencies (solid spheres) and inverse cosine relation (red solid lines).

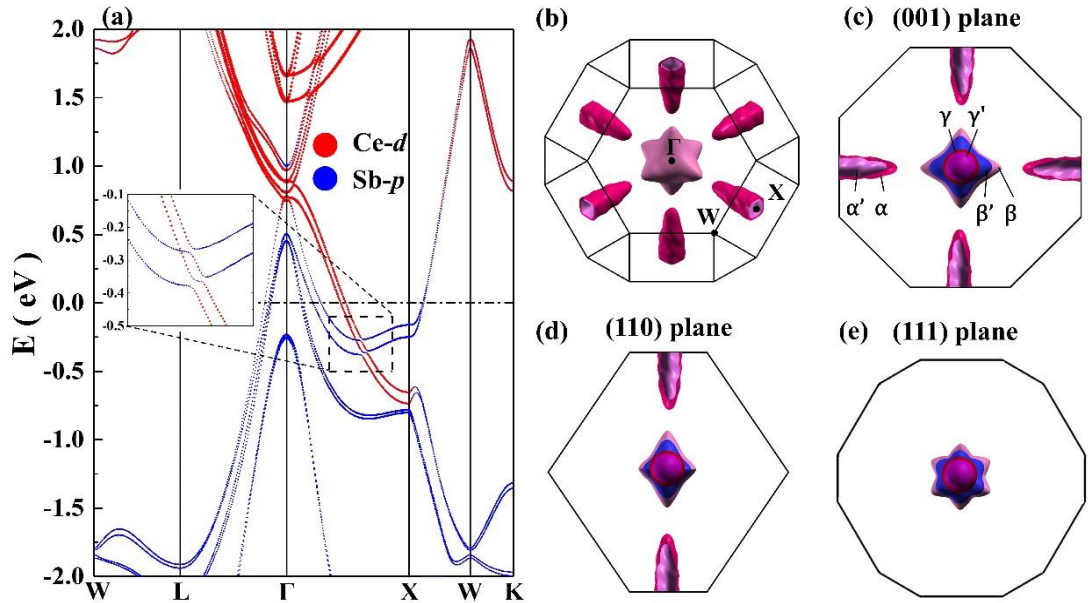

FIG. 5 (a) The projected electronic band structure and (b) Fermi surface of CeSb in the fully polarized ferromagnetic state, which is composed of two electron pockets (α, α′) and four hole pockets (β, β′, γ, γ′). The red and blue dots denote the contribution of Ce-*d* and Sb-*p* bands, respectively. (c)-(e) The Fermi surface projection in (001), (110) and (111) plane, respectively.